\newif\ifsubmit
\submitfalse
\ifsubmit
  \documentclass[sigconf]{acmart}
\else
  \documentclass[twocolumn]{article}
\fi

\usepackage{amsmath}
\usepackage{amssymb}
\usepackage{wasysym}
\usepackage{graphicx}
\usepackage{xspace}
\usepackage{algorithm}
\usepackage{algorithmicx}
\usepackage{algpseudocode}
\usepackage{color}
\usepackage{amsthm}
\usepackage{units}
\usepackage{hyperref}
\usepackage{mathrsfs}

\usepackage{subcaption}

\usepackage{booktabs} % For formal tables

\newcommand{\join}{\bowtie}
\newcommand{\RJ}{ReJOIN\xspace}
\newcommand{\actions}{\mathscr{A}}

\newcommand{\cut}[1] {}
\newcommand{\CM}{ \mathscr{M} }
\newcommand{\PG}{PostgreSQL\xspace}

\ifsubmit
\newcommand{\pg}[1]{{\noindent \bf #1 \ }}
\else
\newcommand{\pg}[1]{\paragraph{#1}}
\fi
\ifsubmit
\setcopyright{none}
\acmConference[aiDM'18]{First International Workshop on Exploiting Artificial Intelligence Techniques for Data Management}{June 2018}{Houston, Texas USA}
\acmYear{2018}
\copyrightyear{2018}

\fi

\begin{document}
\title{Deep Reinforcement Learning for Join Order Enumeration}
%\subtitle{4-Page Paper}

\ifsubmit
\author{Ryan Marcus}
\affiliation{%
  \institution{Brandeis University}
}
\email{ryan@cs.brandeis.edu}

\author{Olga Papaemmanouil}
\affiliation{%
  \institution{Brandeis University}
}
\email{olga@cs.brandeis.edu}
\else
\author{
  Ryan Marcus\\
  Brandeis University\\
  \texttt{ryan@cs.brandeis.edu}
  \and
  Olga Papaemmanouil\\
  Brandeis University\\
  \texttt{olga@cs.brandeis.edu}
}
\date{}
\maketitle
\fi

%\renewcommand{\shortauthors}{Marcus and Papaemmanouil}

\begin{abstract}
Join order selection plays a significant role in query performance.  However, modern query optimizers typically employ static join enumeration algorithms that do not receive any feedback about the quality of the resulting plan. Hence, optimizers often repeatedly choose the same bad plan, as they do not have a mechanism for ``learning from their mistakes''. In this paper, we argue that existing deep reinforcement learning techniques can be applied to address this challenge. These techniques, powered by artificial neural networks, can \emph{automatically} improve decision making by incorporating feedback from their successes and failures. Towards this goal, we present \emph{\RJ}, a proof-of-concept join enumerator, and present preliminary results indicating that \RJ can match or outperform the \PG optimizer in terms of plan quality and join enumeration efficiency.
\end{abstract}

\ifsubmit

\fi

\ifsubmit

%\ccsdesc[500]{Information systems~Query optimization}
%\keywords{Query optimization, join ordering, deep learning}

\maketitle
\fi

\section{Introduction}

Identifying good join orderings for relational queries is one of the most well-known and well-studied problems in database systems (e.g.,~\cite{systemr, volcano, joe_complexity, robust_qo})  since the selected join ordering can have a drastic impact on query performance~\cite{howgood}.  One of the challenges in join ordering selection is \emph{enumerating} the set of candidate orderings and identifying the most cost-effective one.  Here, searching a larger candidate space increases the odds of finding a low-cost ordering, at the cost of spending more time on query optimization. Join order enumerators thus seek to simultaneously minimize the number of plans enumerated and the final cost of the chosen plan. 

Traditional database engines employ a variety of join enumeration strategies. For example, System R~\cite{systemr} uses dynamic programming to find the left-deep join tree with the lowest cost, while Postgres~\cite{url-postgres} greedily selects low-cost pairs of relations until a tree is built. Many commercial products (e.g.,~\cite{url-sqlserver}) include an exhaustive enumeration approach, but allow a DBA to controls the size of the candidate plan set by constraining it structurally (e.g., left-deep plans only), or cutting off enumeration after some elapsed time.

\cut{Early optimizers (e.g., System R~\cite{}) imposed a structural constraint on plans, considering only those whose joins have at least one base relation as input (left-deep enumeration). Given that this sometimes excludes consideration of superior bushy plans~\cite{}, more recent approaches have attempted to use dynamic programming to exhaustively enumerate and evaluate all possible join orderings (e.g., ~\cite{Vertica, PostgreSQL}). In practice, exhaustive enumeration makes optimization of complex queries prohibitively expensive. Thus, many commercial products allow a DBA to set an optimization-level “knob” that controls the size of the candidate plan set. Lowering the optimization level typically constrains this set structurally (e.g., left-deep plans only), or cuts off enumeration after some specified time has passed.}

Unfortunately, these heuristic solutions can often miss good execution plans. More importantly, traditional query optimizers rely on \emph{static} strategies, and hence do not learn from previous experience. Traditional systems plan a query, execute the query plan, and forget they ever optimized this query. Because of the lack of feedback, a query optimizer may select the same bad plan repeatedly, never learning from its previous bad or good choices.

In this paper, we share our vision of a \emph{learning-based} optimizer that  leverages information from previously processed queries, aiming to learn how to optimize future ones  more effectively (i.e., producing better query plans) and efficiently (i.e., spending less time on optimization). We introduce a novel approach to query optimization that is based on \emph{deep reinforcement learning} (DRL)~\cite{deep_rl}, a process by which a machine learns a task through continuous feedback with the help of an artificial neural network. We argue that existing deep reinforcement learning techniques can be leveraged to provide better query plans using less optimization time.  

As a first step towards this goal, we present \emph{\RJ}, a proof-of-concept join order enumerator entirely driven by deep reinforcement learning. In the next section we describe the \RJ  learning framework (Section~\ref{sec:method}) and provide promising preliminary results (Section~\ref{sec:expr}) that show \RJ can outperform PostgreSQL in terms of effectiveness and efficiency of the join enumeration process.

\section{The \RJ Enumerator}\label{sec:method}
Next, we present our proof-of-concept \emph{deep \underline{re}inforcement learning \underline{join} order enumerator}, which we call \RJ.

\begin{figure}
\centering
\includegraphics[width=0.40\textwidth]{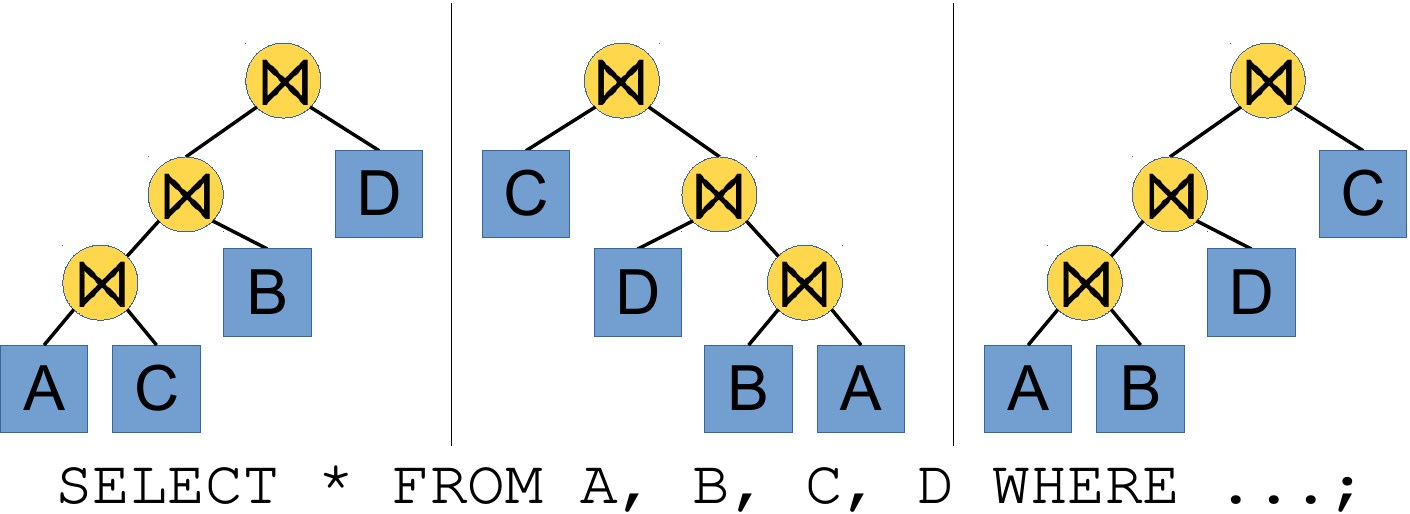}
\caption{\small Three different join orderings}
\label{fig:join_orders}
\vspace{-3mm}
\end{figure}

\pg{Join Enumeration} 
 \RJ assumes a traditional cost-based approach to query optimization used by many modern DBMSs (e.g.,\cite{vertica,url-postgres}). Specifically, given a SQL query as an input, a \emph{join order enumerator} searches a subspace of all possible join orderings and the ``cheapest'' ordering (according to the cost model) is selected for execution. This enumeration does not perform index selection, join operator selection, etc. -- these tasks are left to other components of the DBMS optimizer. 
 A join ordering is captured by a binary tree, in which each leaf node represents a base relation. Figure~\ref{fig:join_orders} shows three possible join trees on the relations $A$, $B$, $C$, and $D$.
 %Both the cost model and the join enumerator may take advantage of cardinality estimations in order to evaluate a query plan or prune part of the search space. 

 %\RJ takes as input a SQL query and Before introducing our proof-of-concept architecture, we first formalize the join enumeration problem and define terminology necessary for a discussion of reinforcement learning.

%Join order enumerators are responsible for selecting a subset of all possible (typically binary) join trees. For example, given a query which joins together the relations $r_1$, $r_2$, and $r_3$, two possible join trees are:

%\begin{align*}
%(r_1 \join r_2) \join r_3 \\
%r_2 \join (r_3 \join r_1)
%\end{align*}

%Depending on the query predicates, relation sizes, and domain cardinality (number of unique values), different join orderings can exhibit drastically different performance. Existing systems 

% Letting the set of all such binary trees be $\mathbb{T}$, join order enumerators search through a (preferably small) subset $T \subset \mathbb{T}$, selecting the join tree $t \in T$ with the lowest cost, as determined by a cost model $\CM$.  

\noindent{\bf Reinforcement Learning} Reinforcement learning assumes~\cite{deep_rl} that an \emph{agent} interacts with an \emph{environment} as follows. The environment tells the agent its current state, $s_t$, and a set of potential actions $\actions_t = \{a_0, a_1, \dots, a_n\}$ that the agent can take. The agent selects an action $a \in \actions_t$, and the environment gives the agent a \emph{reward} $r_t$, with higher rewards being more desirable, along with a new state $s_{t+1}$ and a new action set $A_{t+1}$. This process repeats until the agent selects enough actions that it reaches a \emph{terminal state}, where no more actions are available. This marks the end of an \emph{episode}, after which a new episode begins. The agent's goal is to maximize the reward it receives over episodes by learning from its experience (previous actions, states, and rewards). This is achieved by balancing the \emph{exploration} of new strategies with the \emph{exploitation} of current knowledge.%\cut{\footnote{See \cite{deep_rl} for a more in-depth overview.}}

\noindent{\bf Framework Overview} Next, we formulate the join order enumeration process as a reinforcement learning problem. Each query sent to the optimizer (and to \RJ) represents an episode, and \RJ  learns over multiple episodes (i.e., continuously learning as queries are sent). Each state will represent subtrees of a binary join tree, in addition to information about query join and selection predicates.\footnote{Since predicates  do not change during an episode, we will exclude it from our notation.} Each action will represent combining two subtrees together into a single tree. Note that a subtree can represent either an input relation or a join between subtrees. The episode ends when all input relations are joined (a terminal state). At this point, \RJ assigns a reward to the final output join ordering based on the optimizer's cost model. \cut{ All partial join orderings receive zero reward. \footnote{While traditional optimizers cost each subtree, the \RJ proof-of-concept only costs the final join ordering. This is because (1) it avoids the complication of deeply integrating with the optimizer cost model, as most DBMSes have exposed functionality to cost a complete plan, and (2) our eventual goal is to use the query latency itself as feedback.}}  The final join ordering is then dispatched to the optimizer to perform operator selection, index selection, etc., and the final physical plan is executed by the DBMS.

%The reward given for each action will be zero, except for the action resulting in a terminal state -- this final action, which will result in a join tree $t$, will be  given a reward based on the cost assigned by the cost model $C$ to the final execution plan using the join ordering $t$.  \todo{why? based on the dynamic programming even subplans/ partial plans are costed...}

\begin{figure}
\centering
\includegraphics[width=0.35\textwidth]{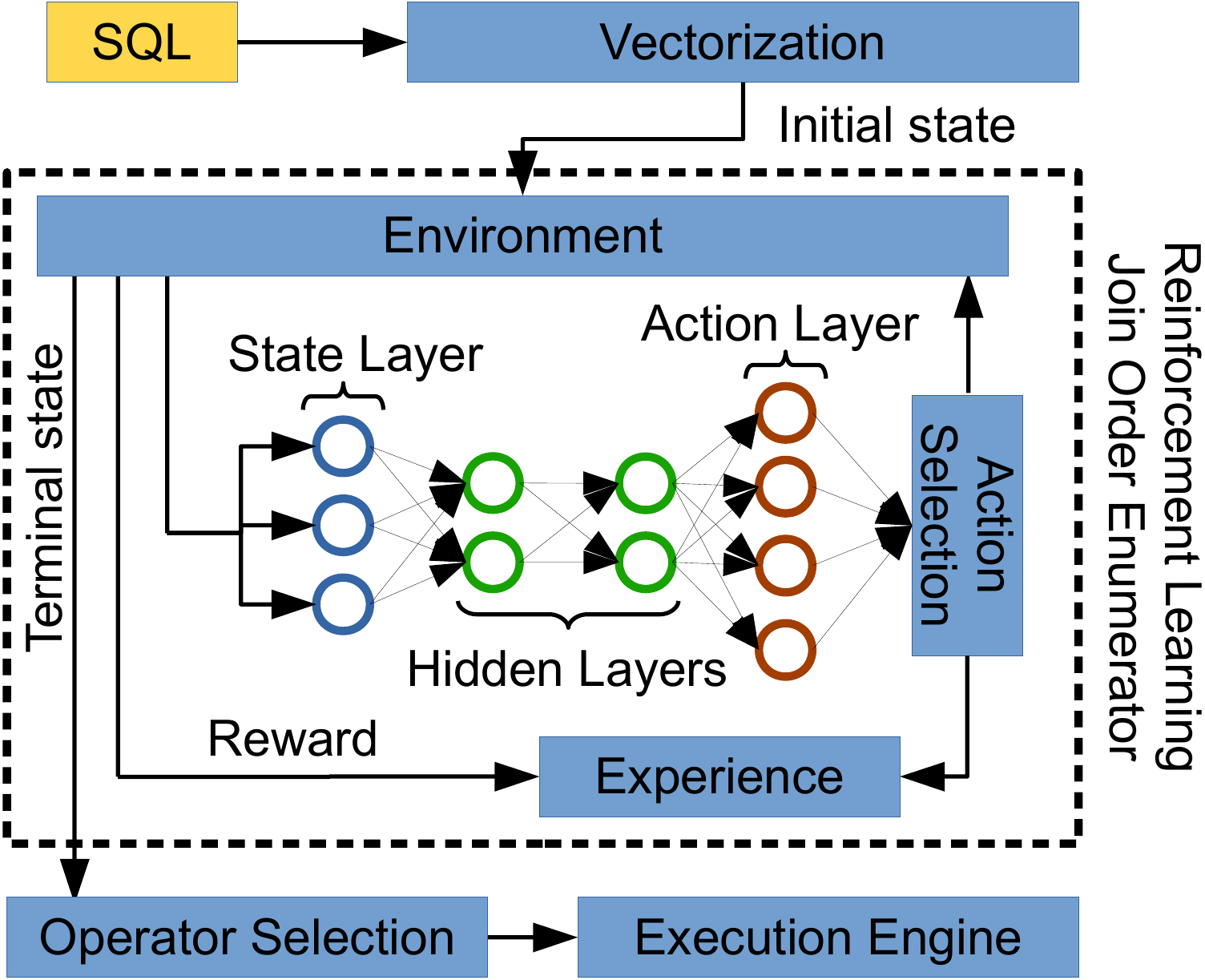}
\caption{\small The \RJ Framework}
\label{fig:system}
\vspace{-5mm}
\end{figure}

The framework of \RJ is shown in Figure~\ref{fig:system}. Formally, given a query $q$ accessing relations $r_1, r_2, \dots, r_n$, we define the initial state of the episode for $q$ as $s_1 = \{r_1, r_2, \dots, r_n\}$. This state is expressed as a \emph{state vector}. This state vector is fed through a neural network~\cite{dnn}, which produces a probability distribution over potential actions. The action set $\actions_i$ for any state is every unique ordered pair of integers from $1$ to $|s_i|$, inclusive: $\actions_i = \left[1, |s_i| \right] \times \left[1, |s_i| \right]$. The action $(x, y) \in \actions_i$ represents joining the $x$th and $y$th elements of $s_i$ together. An action (i.e., a new join) is selected, and sent back to the environment which transitions to a new state.  The state $s_{i+1}$ after selecting the action $(x, y)$ is $s_{i+1} = \left(s_i - \{s_i[x], s_i[y]\}\right) \cup \{s_i[x] \join s_i[y]\}$. The new state is fed into the neural network. The reward for every non-terminal state (a partial ordering) is zero, and the reward for an action arriving at a terminal state $s_f$ (a complete ordering) is the reciprocal of the cost of the join tree $t$, $\CM(t)$, represented by $s_f$, $\frac{1}{\CM(t)}$. Periodically, the agent uses its experience to tweak the weights of the neural network, aiming to earn larger rewards.

%SQL queries are first \emph{vectorized} into a vector representing the state of the \emph{environment} (aka \emph{state vector} described in Section~\ref{s:vectors}). This state vector is fed through a neural network~\cite{dnn}, which produces a probability distribution over potential actions. An action is selected, and sent back to the environment. The environment evaluates the action and produces (1) a reward, which is recorded in the agent's \emph{experience}, and (2) a new state, which is again fed into the neural network. 

\pg{Example} Figure~\ref{fig:actions} shows a potential episode for a query involving four relations: $A$, $B$, $C$, and $D$ The initial state is $s_1 = \{A, B, C, D\}$. The action set $\actions_1$ contains one element for each ordered pair of relations, e.g. $(1, 4) \in \actions_1$ represents joining $A$ with $D$, and $(2, 3) \in \actions_1$ represents joining $B$ with $C$.  The agent chooses the action $(1, 3)$, representing the choice to join $A$ and $C$. The next state is $s_2 = \{ A \join C, B, D \}$. The agent next chooses the action $(2, 3)$, representing the choice to join $B$ and $D$. The next state is $s_3 = \{ A \join C, B \join D\}$. At this point, the agent has only two possible choices, $\actions_3 = \{ (1, 2), (2, 1) \}$. Supposing that the agent selects the action $(1, 2)$, the next state $s_4 = \{ (A \join C) \join (B \join D) \}$ represents a terminal state. At this point, the agent would receive a reward based on the cost model's evaluation of the final join ordering. 
%The second example in Figure~\ref{fig:actions} shows another sequence of actions resulting in the join order $((a \join d) \join c) \join b)$.

\cut{It is straightforward to prove by induction that any binary join tree can be constructed via some sequence of actions, albeit not uniquely (multiple sequences of actions may construct the same tree).}

\subsection{State Vectors}\label{s:vectors}
\cut{One challenge with reinforcement learning is encoding effective information within the state of the environment. In our case,} \RJ uses  a vector representation of each state that captures  information about the join ordering (i.e., the binary tree structure) and the join/selection predicates. Next, we outline a simple vectorization strategy which captures this information and demonstrates that reinforcement learning strategies can be effective \emph{even with simple input data}. %Specifically, we represent each state as three vectors: a tree vector, corresponding to the structure of each partial join ordering (this changes after every action), and two predicate vectors, corresponding to the join and selection predicates. 

\cut{Most off-the-shelf reinforcement learning packages~\cite{tensorforce} require states to be represented as vectors. Intuitively, since the reinforcement learning algorithm will only see the vector representation of each state, and not the states themselves, we wish to encode a sufficient amount of data about each state into the vectorized representation.  Here, we outline a simple vectorization strategy which captures some information about tree structure and join / column predicates. It is by no means an optimal or production-ready vectorization strategy, nor is it intended to be: our goal is to show that reinforcement learning strategies can be effective \emph{even with limited input data.} We represent each state as three vectors: a tree vector, corresponding to the tree structure of each subtree (this changes after every action), and two predicate vectors, corresponding to the join and column predicates (these are fixed for a given episode).}

%\begin{figure}
%\centering
%\includegraphics[width=0.38\textwidth]{figures/vectors-crop.pdf}
%\caption{\small Vectorization of a state}
%\label{fig:vectors}
%\end{figure}

\begin{figure}
\centering
\includegraphics[width=0.48\textwidth]{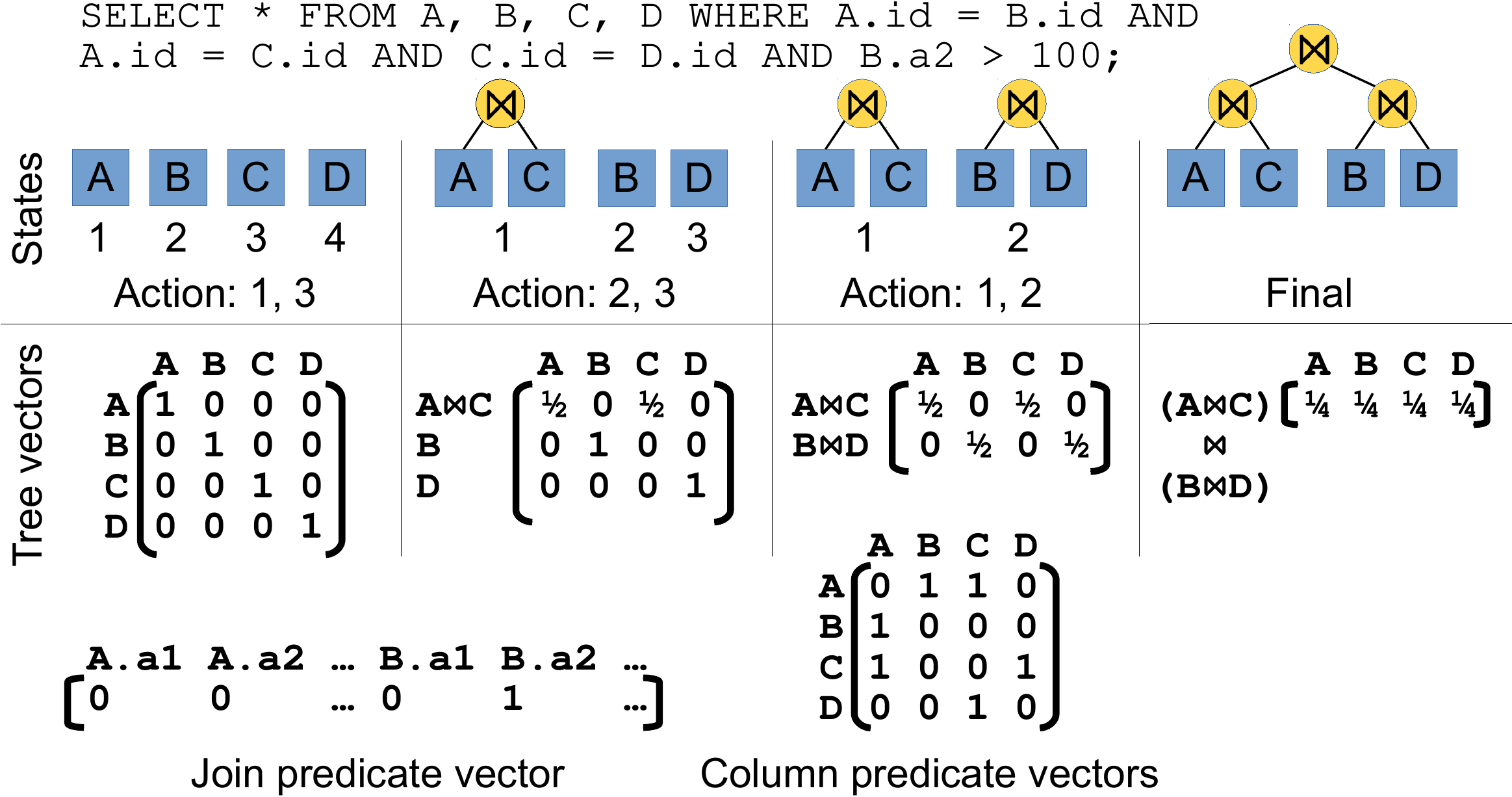}
\caption{\small Two possible join order selection episodes}
\label{fig:actions}
\vspace{-5mm}
\end{figure}

\noindent{\bf Tree Structure} To capture tree structure data, we encode each binary subtree (i.e., join ordering decided so far) $x \in s_j$ as a row vector $v$ of size $n$, where $n$ is the total number of relations in the database. The value $v_i$ is zero if the $i$th relation is not in $x$, and equal to $\frac{1}{h(i, x)}$ otherwise, where $h(i, x)$ is the height of the relation $r_i$ in the subtree $x$ (the distance from the root). In the example in Figure~\ref{fig:actions}, the first row of the tree vector for the second to last state, $\{A \join C, B \join D\}$, corresponds to $(A \join C)$. The third column of the first row has a value of $\frac{1}{2}$, corresponding to $C$ having a height of 2 in the subtree. The second column of the first row has a value of zero since the relation $B$ is not included in the subtree. %Figure~\ref{fig:join_orders} shows more examples of join orders and tree vectors.

\pg{Join Predicates} To capture critical information about join predicates, we create an $n \times n$ binary symmetric matrix $m$ for each episode. The value $m_{i,j}$ is one if there is a join predicate connecting the $i$th and the $j$th relation, and a zero otherwise. This simple representation captures feasible equi-joins operations. Figure~\ref{fig:actions} shows an example of such a matrix $m$. The value $m_{2,1} = m_{1,2} = 1$ because of the predicate \texttt{A.id = B.id}. The value $m_{2,3} = m_{3, 2} = 0$ because there is no join predicate connecting $B$ and $C$. %The diagonal of $m$ is all zeros because there are no self-join predicates.

\cut{does not capture complex join predicates (such as inequalities or more advanced functions), but it does capture information about which relations have no predicates between them (the zeros) -- if two relations have no connecting predicates, it is very unlikely that joining them together directly is a good idea}

\noindent{\bf Selection Predicates} The selection predicate vector is a $k$-dimensional vector, where $k$ is the total number of attributes in the database (the total number of attributes across all relations). The $i$th value is one when the $i$th attribute has a selection predicate in the given query, and zero otherwise. This does reveal which attributes are not used to filter out tuples. For example, in Figure~\ref{fig:actions} the value corresponding to \texttt{B.a2} is one because of the predicate \texttt{B.a2 > 100}. %All other values are zero, because there are no other column predicates.  

%\noindent{\bf Discussion} It bears repeating that this is not a production-ready or optimal vectorization strategy. A significant amount of information (predicate values, histograms, etc.) is not included, but this is intentional: our goal is to demonstrate that reinforcement learning approaches can be effective \emph{even with limited input data.} It is worth noting the potential scalability issues with this vectorization scheme, as the size of the join predicate matrix is $O(n^2)$, with $n$ being the total number of relations in the database, and the column predicate vector scaling with $O(k)$, with $k$ being the number of columns in the database. We are currently investigating more effective and compact vectorization strategies.

\subsection{Reinforcement Learning}

\label{sec:deep_rl}
\cut{In this section we describe the deep learning approach used by \RJ.}
\pg{Policy gradient} Our framework relies on policy gradient methods~\cite{reinforce}, one particular subset of reinforcement learning. Policy gradient reinforcement learning agents select actions based on a parameterized \emph{policy} $\pi_\theta$, where $\theta$ is a vector that represents the policy parameters. Given a state $s_t$ and an action set $\actions_t$, the policy $\pi_\theta$ outputs a score for each action in $\actions_t$ (in our context, a score for combining two join subtrees). Actions are then selected using various methods~\cite{deep_rl}.\cut{Formally:
\begin{equation*}
\pi_\theta : (s_t, A_t) \to a_i \in A_t
\end{equation*}}
%To balance exploration and exploitation, the scores are typically modeled as a probability distribution.
 
Reinforcement learning aims to optimize the policy $\pi_\theta$ over episodes, i.e., to identify the policy parameters $\theta$ that optimizes the expected reward $J_\pi(\theta)$.  However,  the reward $J_\pi(\theta)$ is typically not feasible to precisely compute and hence policy gradient methods search for the optimal policy parameters $\theta$ by constructing an estimator $E$ of the \emph{gradient} of the reward: $E(\theta) \approx \nabla_\theta J_\pi(\theta)$. 

\cut{For real-world applications require that any change to the policy parameterization has to be smooth as drastic changes can be hazardous for the system as well as useful initializations of the policy based on domain knowledge would otherwise vanish after a single step. 
For these reasons,} Given an estimate $E$, \emph{gradient ascent} methods  tune the initial random parameters $\theta$ by incrementing  each parameter in $\theta_i$ by a small value when the gradient $\nabla_{\theta_i} J_\pi(\theta)$ is positive (the positive gradient indicates that a larger value of $\theta_i$ will increase the reward), and decrementing the parameters in $\theta_i$ by a small value when the gradient is negative. 

\cut{The performance function, or the expected reward that a policy will receive per episode, is denoted $J_\pi(\theta)$. A reinforcement learning agent thus seeks the vector $\theta$ that maximizes the reward $J_\pi(\theta)$, but the reward $J_\pi(\theta)$ is typically not feasible to precisely compute. Policy gradient methods search for such a vector $\theta$ by constructing an estimator $E$ of the \emph{gradient} of the performance function: $E(\theta) \approx \nabla_\theta J_\pi(\theta)$. Given an estimate $E$, \emph{gradient ascent} methods can be used to tune an initial random $\theta$. Intuitively, these methods work by incrementing $\theta_i$ by a small value when the gradient $\nabla_{\theta_i} J_\pi(\theta)$ is positive (the positive gradient indicates that a larger value of $\theta_i$ will increase performance), and decrementing $\theta_i$ by a small value when the gradient is negative.
}

\pg{Policy gradient deep learning} {Policy gradient deep learning methods} (e.g., ~\cite{ppo}) represent the policy $\pi_\theta$ as a neural network, where $\theta$ is the network weights, thus enabling the efficient differentiation of $\pi_\theta$~\cite{dnn}. Figure~\ref{fig:system} shows the policy network we used in \RJ. A vectorized representation of the current state is fed into the \emph{state layer}, where each value is transformed and sent to the first hidden layer. The first hidden layer transforms and passes its data to the second hidden layer, which passes data to the final \emph{action layer}. Each neuron in the action layer represents one potential action, and their outputs are normalized to form a probability distribution. 
The policy $\pi_\theta(s_i, A_i)$ selects actions by sampling from this probability distribution, \cut{Selecting the \emph{mode} of the distribution instead of sampling from the distribution would represent a \emph{pure exploitation} strategy. Choosing an action uniformally at \emph{random} would represent a \emph{pure exploration} strategy.} which balances exploration and exploitation~\cite{reinforce}.

The policy gradient $\nabla_\theta J_\pi(\theta)$ is  estimated using samples of previous episodes (queries). Each time an episode is completed (a join ordering for a given query is selected), the \RJ agent records a new  observation $(\theta, r)$. Here, $\theta$  represents the policy parameters used for that episode and the final cost (reward) $r$ received. Given a set of experiences over multiple episodes $X = \{(\theta^0, r^0), (\theta^1, r^1), \dots, (\theta^2, r^2)\}$, various advanced techniques can be used to estimate the gradient $E(\theta)$ of the expected reward~\cite{ac,ppo}.

\cut{While relatively simple, deep reinforcement learning techniques have demonstrated success in advanced tasks such as playing Atari video games and controlling complex robots~\cite{deep_rl}.}

\section{Preliminary Results}
\label{sec:expr}

Here, we present  preliminary experiments that indicate that \RJ can generate join ordering with cost and latency as good  (and often better) as the ones generated from the \PG~\cite{url-postgres} optimizer.

\begin{figure}
\centering
\includegraphics[width=0.45\textwidth]{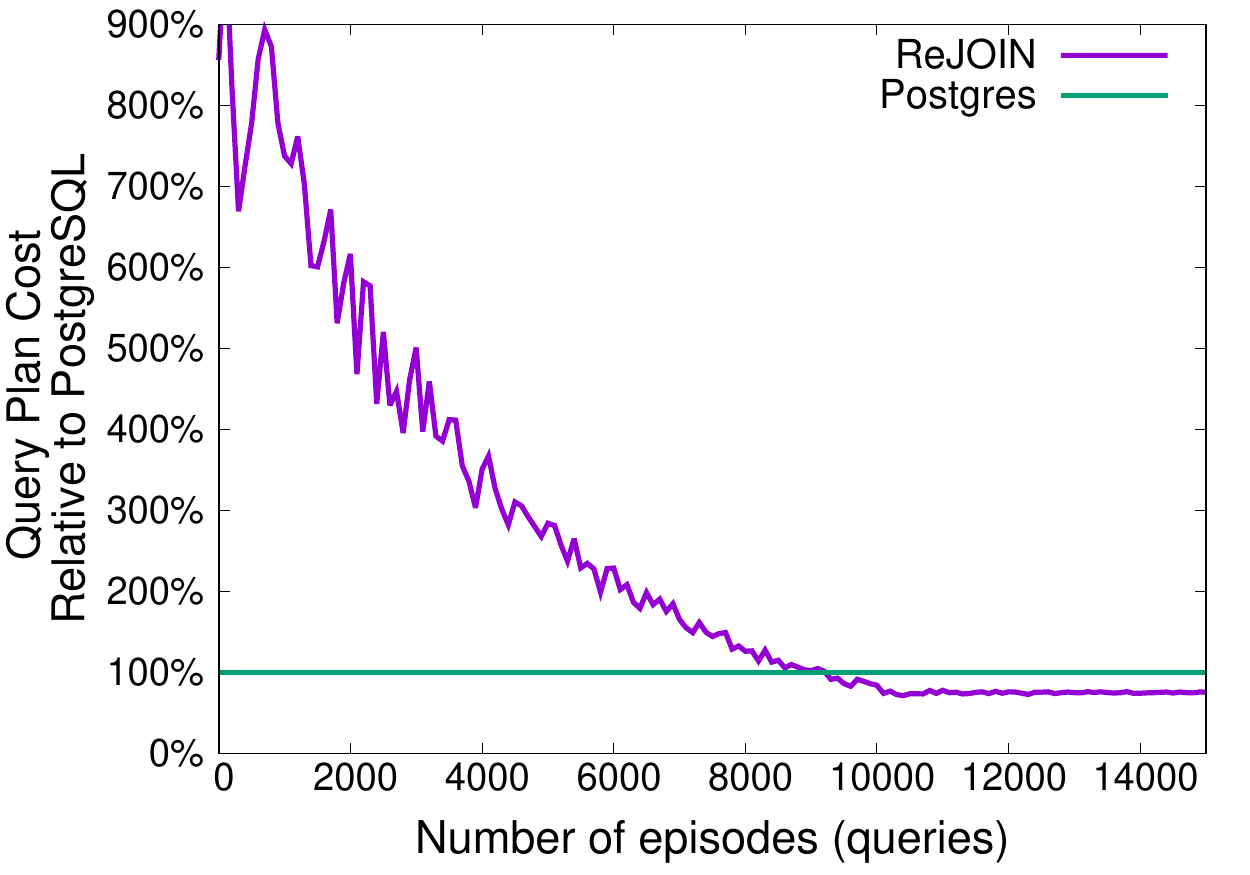}
\caption{\small \RJ Convergence}
\vspace{-3mm}
\label{fig:converge}
\end{figure}

\begin{figure*}
\centering
\begin{subfigure}{0.32\textwidth}
	\centering
	\includegraphics[width=\textwidth]{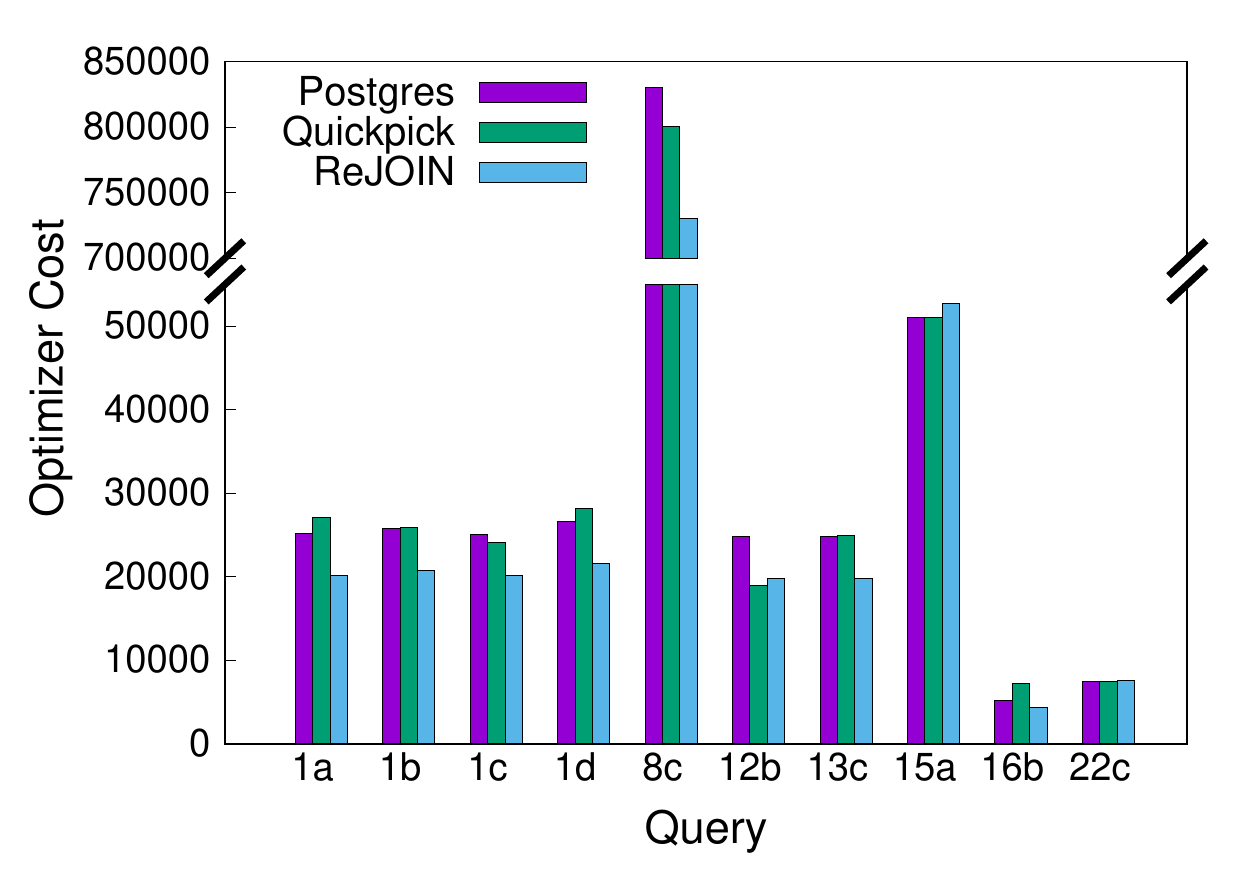}
	\caption{\small Cost of join orderings}
	\label{fig:cost}
\end{subfigure}
\begin{subfigure}{0.32\textwidth}
	\includegraphics[width=\textwidth]{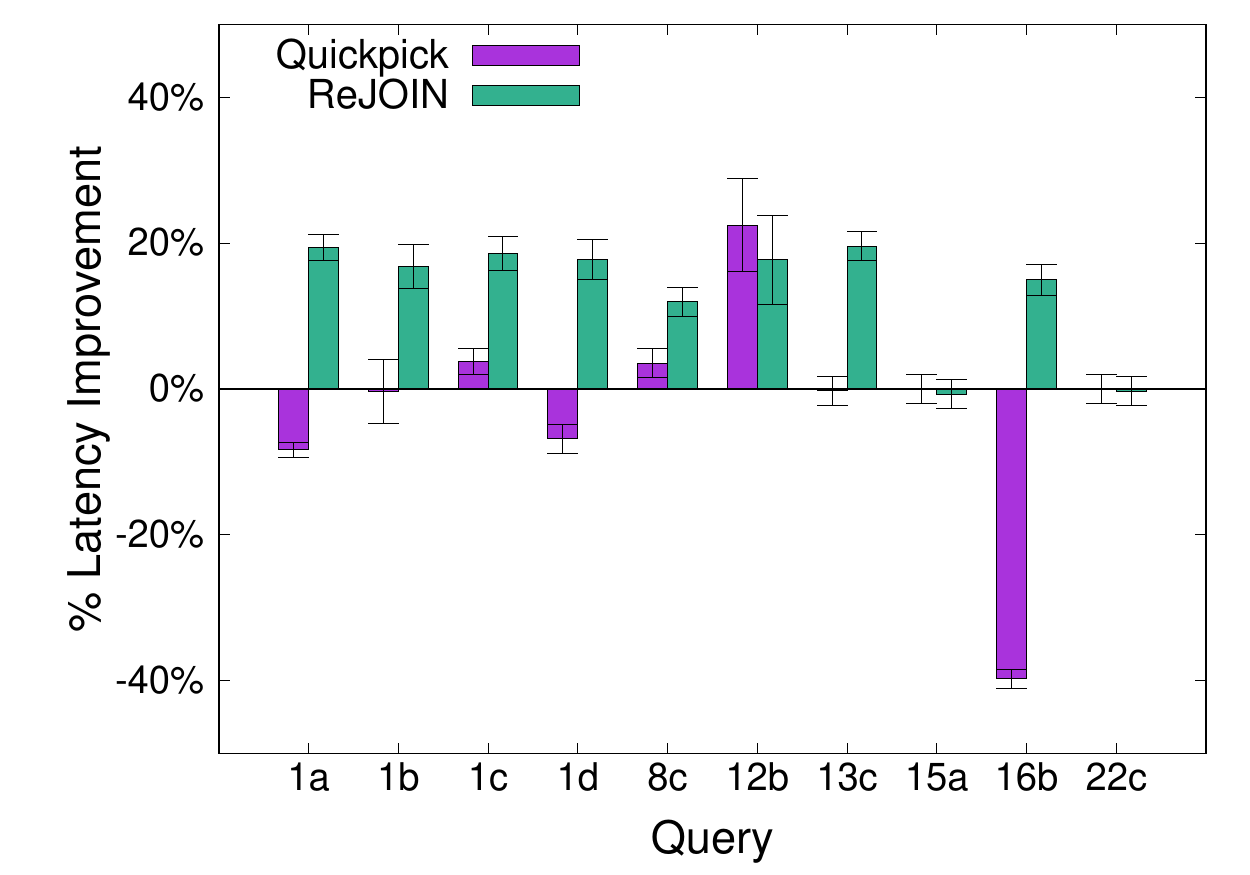}
	\caption{\small Latency of generated plans}
	\label{fig:latency}
\end{subfigure}
\begin{subfigure}{0.32\textwidth}
	\includegraphics[width=\textwidth]{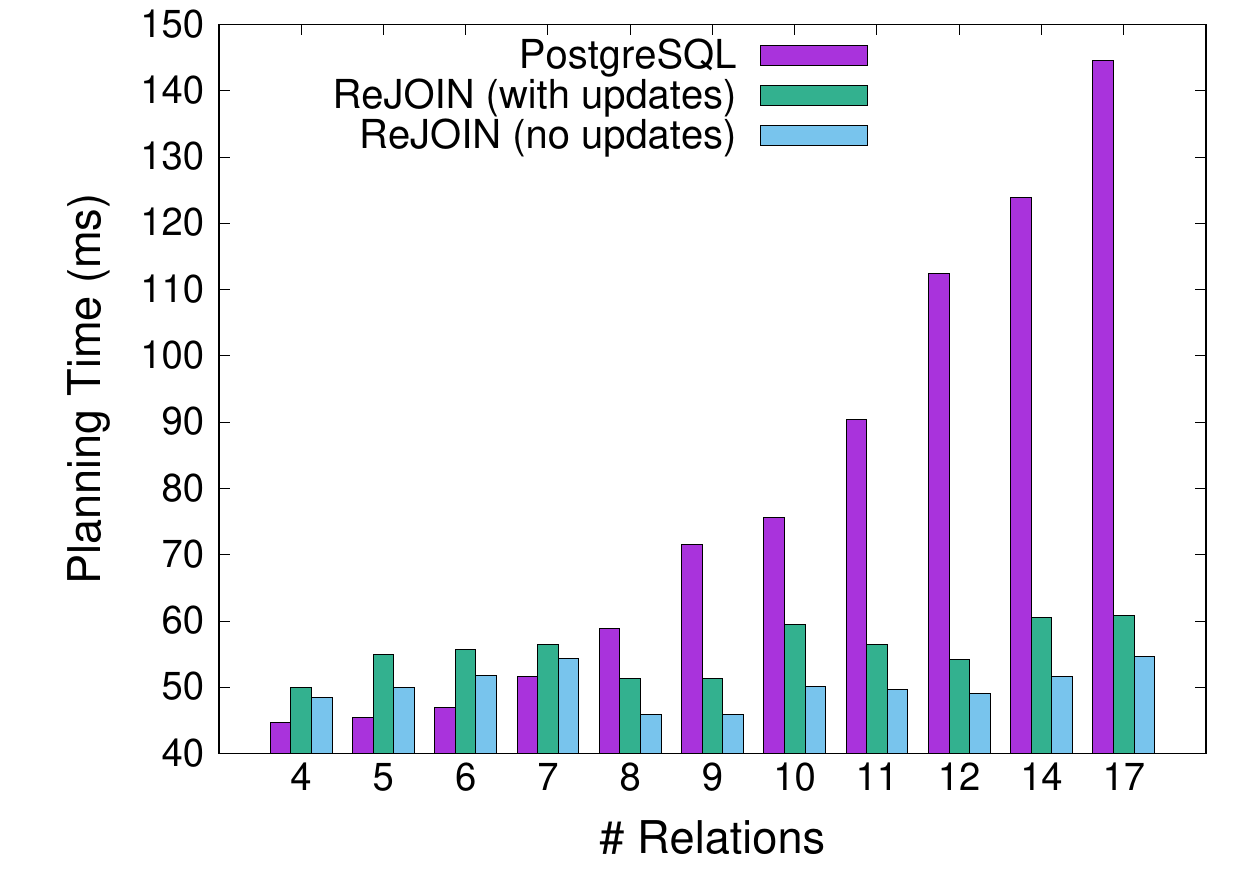}
	\caption{\small Optimization time}
	\label{fig:plan}
\end{subfigure}
\vspace{-3mm}
\caption{\small Effectiveness and efficiency results}
\vspace{-3mm}
\label{fig:variances}
\end{figure*}

%\subsection{Experimental Setup}

Our experiments are based on the Join Order Benchmark (JOB), a set of queries used in previous assessments of query optimizers~\cite{howgood}. The benchmark includes a set of 113 query instances of 33 query templates over the freely-available IMDB dataset. We have created a virtual machine pre-loaded with the dataset~\cite{url-imdbgithub}. Each query joins between 4 and 17 relations, and the two largest relations contain 36M and 15M rows.
\RJ is trained on 103 queries and tested on 10 queries. Our testing query set includes all instances of one randomly selected query template (template \#1), in addition to six other randomly selected queries. %The reinforcement learning agent never trains on these ten queries, and is trained only on the remaining 103 queries.

The total database size was 11GB (all primary and foreign keys are indexed) using PostgreSQL~\cite{url-postgres} on a virtual machine with 2 cores, 4GB of RAM and a maximum shared buffer pool size of 1GB. We configured PostgreSQL to execute the join ordering generated by \RJ instead of using its own join enumerator~\cite{url-pgeo}.  %The virtual machine had 2 cores, and was running on a server with an Intel Xeon E5-2640v4 CPU. 

\RJ uses the  \emph{Proximal Policy Optimization} (PPO) algorithm~\cite{ppo}, an off-the-shelf~\cite{tensorforce} state-of-the-art DRL technique.  We used two hidden layers with 128 rectified linear units (ReLUs)~\cite{relu} each.

%Our prototype implementation worked by processing a query and building a join ordering with \RJ, and then passing that join ordering to Postgres using explicit join clauses. This forces Postgres to execute the  join ordering generated by \RJ, bypassing the one generated by its own join enumeration process.\footnote{Additionally, the \texttt{geqo\_threshold} and \texttt{join\_collapse\_limit} variables must be appropriately set.} Given a join ordering, the Postgres optimizer then selects operators, indexes, etc. before processing the query.

%\noindent {\bf Learning Techniques} We evaluated several deep reinforcement learning algorithms, including DQN~\cite{dqn}, A3C~\cite{a3c}, TRPO~\cite{trpo}, and Double DQN~\cite{ddqn} through the TensorForce~\cite{tensorforce} library. After tuning each, 

%\cut{\noindent {\bf Evaluation Metrics} We evaluate \RJ in terms of the Postgres cost model, actual runtime latency, and query planning time. }

\pg{Learning Convergence} To evaluate its learning convergence, we ran the \RJ algorithm repeatedly, selecting a random query from the training set at the start of each episode. The results are shown in Figure~\ref{fig:converge}. The x-axis shows the number of queries (episodes) the \RJ agent has learned from so far, and the y-axis shows the cost of the generated plan relative to the plan generated by the \PG optimizer, e.g. a value of 200\% represents a plan with double the cost of the plan selected by the \PG optimizer. \RJ starts with no information, and thus initially perform very poorly. As the number of observed episodes (queries) increases, the performance of the algorithm improves. At around 8,000 observed queries, \RJ begins to find plans with lower predicted cost than the \PG optimizer. After 10,000 queries, {the average cost of a plan generated by \RJ is 80\% of the cost of a plan generated by \PG}. \emph{This demonstrates that, with enough training, \RJ can learn to produce effective join orderings.}

\pg{Join Enumeration Effectiveness}
To evaluate the effectiveness of the join orderings produced by \RJ, we first trained our system over 10,000 queries randomly selected from our 103 training queries (the process took about 3 hours). Then, we used the generated model to produce a join ordering for our 10 test queries. For each test query, we used the converged model to generate a join ordering, but we did not update the model (i.e., the model did not get to add any information about the test queries to its experience set). We recorded the cost (according to the PostgreSQL cost model) of the plans resulting from these test join orderings as well as their execution times. We compare  the effectiveness of  \RJ with PostgreSQL as well as \emph{Quickpick}~\cite{quickpick}, which heuristically samples 100 semi-random join orderings and selects the join ordering which, when given to the DBMS cost model, results in the lowest-cost plan.  

\underline{Optimizer costs} We first evaluated the join orderings produced by the  \RJ enumerator based on the cost model's assessment. The costs of the plans generated by \PG's default enumeration process, Quickpick, and \RJ on the 10 test queries are shown in Figure~\ref{fig:cost}. ``Query XY'' on the x-axis refers to the instance Y of the template X in the JOB benchmark.  On average, \emph{\RJ produced join orderings that resulted in query plans that were 20\% cheaper than the PostgreSQL optimizer.} In the worst case, \RJ produced a cost only 2\% higher than the PostgreSQL optimizer (Query 15a). \emph{This shows that \RJ was able to learn a generalizable join order enumeration strategy which outperforms or matches the cost of the join ordering produced by the PostgreSQL optimizer.} The relatively poorer performance of Quickpick  demonstrates that \RJ's good performance is not due to random chance. % -- if it were the case that most join orderings performed as well as or better than the one selected by PostgreSQL, Quickpick's performance would demonstrate this \todo{still not clear why/how? this sentence said really nothing new}.

\underline{Query latency} \cut{We additionally checked the execution times of the resulting plans to verify that \RJ's plans had a lower execution time (not just a lower cost according to the cost model).} Figure~\ref{fig:latency} shows the latency of the executed query plans created by Quickpick and \RJ relative to the performance of the plan selected by the PostgreSQL optimizer. Here, each test query is executed 10 times with a cold cache. The graph shows the minimum, maximum, and median latency improvement. \emph{In every case, the plans produced by \RJ's join ordering outperform or matches the plan produced by PostgreSQL.} Hence, \RJ can produce plans with a lower execution time (not just a lower cost according to the cost model).  Again, the relatively poorer performance of Quickpick demonstrates that \RJ is not simply guessing a random join ordering.

\pg{Join Enumeration Efficiency}
A commonly-voiced opinion about neural networks -- and machine learning in general -- is that they require operations that are too expensive to include in database internals. Here, we demonstrate that approaches like \RJ can actually \emph{decrease} query planning time. Figure~\ref{fig:plan} shows the average total query optimization time for the 103 queries in the training set, grouped by the number of relations to be joined. We include the planning time for \RJ with and without policy updates. 

\underline{Planning latency} For PostgreSQL,  as expected, more relations in the query resulted in higher optimization time, as more join orderings need to be considered and ran through the cost model. However, \RJ can apply its model in time linear to the number of relations: at each round, two subtrees are joined until a complete join ordering is produced. As a result, \emph{while \PG's query optimization time increases worse-than-linearly, the query optimization time for \RJ is (relatively) flat.}

\underline{Policy update overhead} The additional overhead of performing policy updates per-episode using PPO is relatively small, e.g. $< 12ms$. However, once the \RJ model is sufficiently converged, \emph{policy updates can be skipped to reduce query planning times by an additional 10\% to 30\%, achieving even shorter query planning times.}

%\emph{In our experiments, {using \RJ to fix the join ordering resulted in a 10\% to 90\% decrease in total query planning time, compared to the PostgreSQL join enumerator.}}

\section{Open challenges \& ongoing work}
\label{sec:conclusion}
Our simple reinforcement learning approach to join enumeration indicates that there is room for advancement in the space of applying deep reinforcement learning algorithms to query optimization problems. Overall, we believe the \RJ opens up exciting new research paths, some of which we highlight next.

\pg{Latency optimization} Cost models depend on cardinality estimates, which are often error-prone. It would be desirable to use the actual latency of an execution plan, as opposed to a cost model's estimation, as a reward signal. \RJ uses the cost model as a proxy for query performance because it enables us to quickly train the algorithm for a large number of episodes -- executing query plans, especially the poor ones generated in early episodes, would be overly time-consuming. We are currently investigating techniques~\cite{dqfd} to ``bootstrap'' the learning process by first observing an expert system (e.g., the Postgres query optimizer), mimicking it, and then improving on the mimicked strategy.

\pg{End-to-end optimization} \RJ only handles join order selection, and requires an optimizer to select operators, choose indexes, coalesce predicates, etc. One could begin expanding \RJ to handle these concerns by modifying the action space to include operator-level decisions.

%\clearpage
\ifsubmit
\bibliographystyle{acm}
\bibliography{ryan-cites-short}
\else
\onecolumn
\bibliographystyle{abbrv}
\bibliography{ryan-cites-long}
\fi
\end{document}